\documentclass[10pt,twocolumn,letterpaper]{article}

\usepackage[pagenumbers]{cvpr} 

\definecolor{cvprblue}{rgb}{0.21,0.49,0.74}
\usepackage[pagebackref,breaklinks,colorlinks,allcolors=cvprblue]{hyperref}

\def\confName{CVPR}
\def\confYear{2026}

\title{Real-Time Speech Enhancement via a Hybrid ViT: A Dual-Input Acoustic-Image Feature Fusion}

\author{Behnaz Bahmei\\
Mechatronic Systems Engineering\\
Simon Fraser University\\
bbahmei@sfu.ca
\and
Siamak Arzanpour\\
Mechatronic Systems Engineering\\
Simon Fraser University\\
arzanpour@sfu.ca
\and
Elina Birmingham\\
Faculty of Education\\
Simon Fraser University\\
ebirmingham@sfu.ca}

\begin{document}
\maketitle
\begin{abstract}
Speech quality and intelligibility are significantly degraded in noisy environments. This paper presents a novel transformer-based learning framework to address the single-channel noise suppression problem for real-time applications. Although existing deep learning networks have shown remarkable improvements in handling stationary noise, their performance often diminishes in real-world environments characterized by non-stationary noise (e.g., dog barking, baby crying). The proposed dual-input acoustic-image feature fusion using a hybrid ViT framework effectively models both temporal and spectral dependencies in noisy signals. Designed for real-world audio environments, the proposed framework is computationally lightweight and suitable for implementation on embedded devices. To evaluate its effectiveness, four standard and commonly used quality measurements, namely PESQ, STOI, Seg SNR, and LLR, are utilized. Experimental results obtained using the Librispeech dataset as the clean speech source and the UrbanSound8K and Google Audioset datasets as the noise sources, demonstrate that the proposed method significantly improves noise reduction, speech intelligibility, and perceptual quality compared to the noisy input signal, achieving performance close to the clean reference.
\end{abstract}  
\section{Introduction}
\label{sec:intro}

The presence of noise affects the perceptual quality of speech signals and decreases the performance of speech-related applications such as automatic speech recognition (ASR) ~\cite{zhang2023noise}, hearing aid devices ~\cite{petersen2022hearing}, and audio/video communications. It can even be distressing for some individuals, such as those on the autism spectrum disorder (ASD) with auditory hypersensitivity. Noise suppression or speech enhancement is widely used as an effective solution for converting a noisy signal to a high-quality and clean signal. Noise suppression algorithms aimed to attenuate the ambient noise components from speech signals to make them more intelligible. 

Over the decades, several noise suppression methods and systems have been proposed in the literature.  Traditional methods  ~\cite{berouti1979enhancement, reddy2007softmask, grais2011nmf, kamath2002multiband, scalart1996apriori, roy2022kalman} employed different signal processing analysis methods, such as signal-to-noise ratio (SNR), Wiener filtering, and spectral subtraction algorithm on noise structures to overcome this problem. However, they are not effective for complex, non-stationary, and highly dynamic noises like sirens and dog barking, sounds that are commonly found in real-world environments.

In recent years, Deep Learning (DL) methods such as deep neural networks (DNNs) ~\cite{xu2015regression, yu2020kalman}, convolutional neural networks (CNNs) ~\cite{karthik2023cnn}, long short-term memory (LSTM) ~\cite{fernandez2020intelligibility}, recurrent neural networks (RNNs) ~\cite{hu2020dccrn, wenhao2021gcrn}, and deep denoising autoencoders (DDAE) ~\cite{lu2013denoising} have achieved remarkable performance in speech enhancement and audio denoising. These algorithms address the issues in traditional methods by proposing several single and multi-channel approaches with either supervised or unsupervised learning models. The superior performance of DL models, among other data-driven models, is attributed to their exceptional nonlinear mapping ability, which directly transforms the noisy signal to the clean one. DL models employ two approaches, including (i) directly predicting the clean signal ~\cite{germain2019featureloss} and (ii) predicting a mask for filtering ~\cite{williamson2017complexmask}.  Despite the success of DL algorithms in offline speech enhancement, their performance for real-time audio processing applications and complex noises such as sirens, dog barking, car horns, and constructional noises is not well investigated. In some cases, the speech is distorted and not intelligible, or there are still residual noises in the clean signal. For adoption in a real-time application, a DL based system needs to complete several steps, including pre-processing, feature extraction, prediction, and post-processing in a short timeframe such that its output is intelligible and without any noticeable delay by the human listener (typically below 40 ms ~\cite{stone1999delay}).

This makes noise suppression for speech enhancement a highly complex problem, and as a result, achieving satisfactory performance with deep learning models remains challenging. In recent years, Vision Transformers (ViT) have become the dominant paradigm in computer vision due to their superior capability in capturing long-range spatial dependencies and contextual modeling from image data ~\cite{dosovitskiy2021vit}. Inspired by this unparalleled success, we seek to perform an architectural power transfer to the domain of signal denoising. In this work, we introduce a modified Vision Transformer architecture optimized for extracting rich features from 2D acoustic representations. Specifically, our model utilizes both the raw audio signal (to preserve phase information) and the spectrogram, which serves as a frequency-temporal image. This dual-data input allows us to effectively segment the spectrogram into visual tokens and leverage the ViT's attention mechanism to accurately extract and model the complex spatiotemporal and frequency-domain relationships crucial for targeted speech enhancement. This approach represents a significant step towards adapting machine vision architectures to solve image-like signal processing challenges.

This paper proposes a novel learning framework based on a Vision Transformer encoder to address the single-channel noise suppression problem for real-time applications. The framework processes audio files comprising a combination of speech and noise captured through a microphone. By extracting and analyzing features from the noisy signal, the model constructs a ratio mask to suppress noise and enhance speech quality. Unlike previous approaches, the proposed method incorporates both spectral and raw audio features, improving its ability to handle complex and dynamic noise environments. The framework includes pre-processing and post-processing steps to ensure high-quality, intelligible output while maintaining real-time performance with minimal latency. The framework is tested on challenging noise structures, such as dog barking, sirens, vacuum cleaners, and engine noise, demonstrating robust performance in real-world scenarios.

To summarize, the main contributions of this paper are as follows:
\begin{enumerate}
    \item An adapted Vision Transformer architecture is proposed and tailored for processing 2D acoustic image data (spectrogram). 

    \item A robust input pipeline utilizing both the spectrogram image and the raw audio signal is proposed to ensure superior noise attenuation capabilities across stationary, non-stationary, and highly dynamic noise environments. 

    \item The system is designed with a lightweight architecture, enabling it to run in real-time with minimal computational overhead. 
\end{enumerate}

The composition of the paper is as follows. In Section 2, the materials and methods, including pre-processing, model structure, post-processing and experimental settings, are discussed. Section 3 presents the results in detail, and the conclusion and future directions are presented in Section 4.

\section{Materials and Methods}
\label{sec:formatting}

In this section, the procedure of the noise suppression algorithm is explained. First, in the pre-processing part, the features of noisy and clean signals are extracted as network input and output. Then, the architecture of the model is explained in detail. Following that, the process of constructing the clean signal is described in the post-processing part. Finally, the overall learning framework, the network training structure, and real-time processing are introduced in the experimental setup and real-time processing parts, respectively.

\subsection{Pre-processing}

Figure 1 shows the overall block diagram for the entire system, including the pre-processing, the model, and the post-processing.
According to the diagram, a raw noisy signal is taken as the input, and a clean signal is produced as the output. The noisy signal is generated using a combination of a clean signal and a noise signal, which are assumed to be uncorrelated and can be formulated as,

\begin{equation}
X(k) = S(k) + G * N(k)
  \label{eq:important}
\end{equation}

where X(k) is the noisy mixture, S(k) is the clean signal, N(k) is the noise signal, G is the noise gain, and k is the discrete-time index. The clean signal could be a combination of pleasant sounds such as speech, home or office background, birds chirping, and so on. The noise gain is used to create various distributions of noise and clean signals in the mixture, resulting in different signal-to-noise ratios (SNRs) to simulate a real-world noisy environment. 
The aim is to extract the signal S(k) from the noisy signal X(k).

\begin{itemize}
    \item Spectral Branch Pre-Processing
\end{itemize}
The first step of the pre-processing part is Filter 1. This is a bandpass filter that is considered a passive attenuation gate for improving the quality or intelligibility of the mixed signal. This filter allows a frequency range between 40Hz and 4 kHz to pass and cuts off the other frequencies outside this range. This frequency range is selected to suppress out-of-range noises that fall outside the human voice spectrum ~\cite{balasubramaniam2020bandpass}. 

\begin{figure*}
  \centering
  \includegraphics[width=0.9\linewidth]{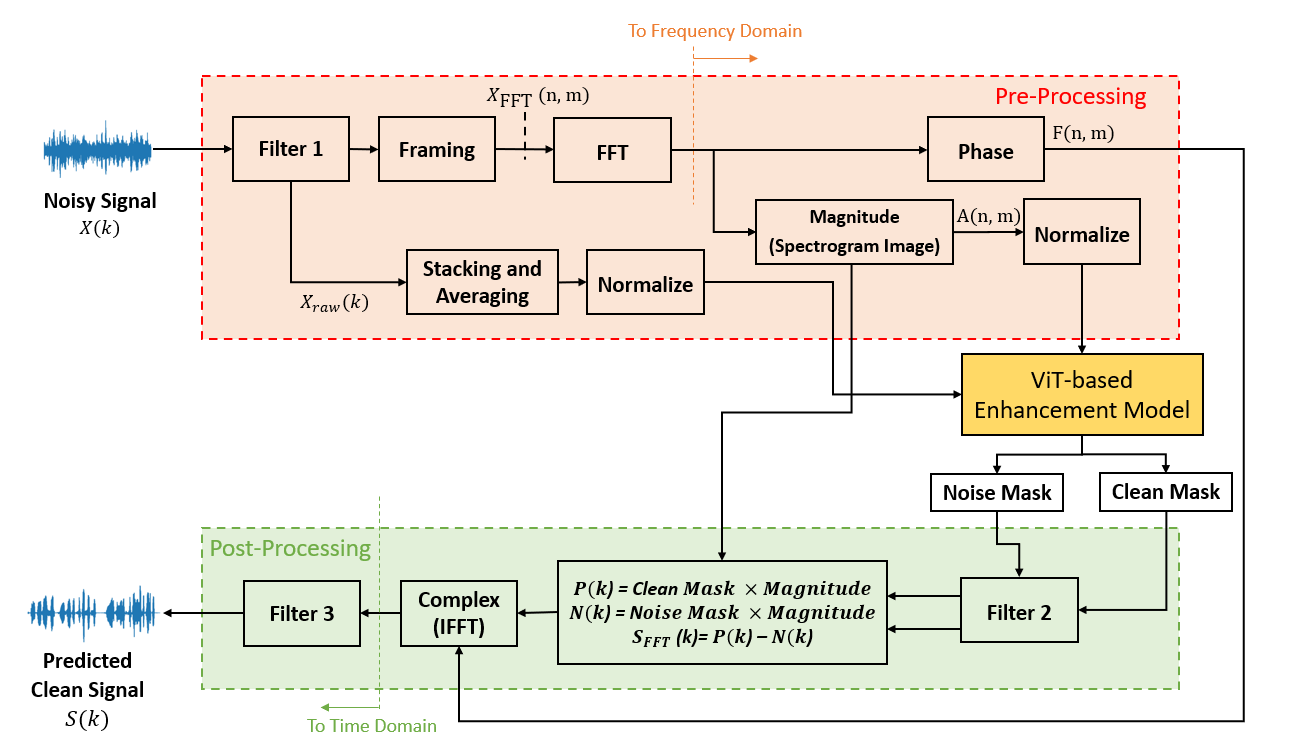}
  \caption{The proposed noise suppression architecture.}
\end{figure*}

The next step in Figure 1 is framing, where the filtered mix signal, X(k), is divided into short-length frames and then a Hanning window with 50\% overlap is applied to minimize spectral distortion and maintain continuity across frames. Afterward, the Short Time Fourier Transform (STFT) is used to convert the framed signal into the frequency domain. This transform produces a complex matrix spectrogram (an acoustic image representation) that is linearly scaled and factored into phase and magnitude components ~\cite{kelkar1983parseval}. $\hat{X}_{FFT}$ represents the STFT of the signal and can be introduced by
\begin{equation}
\hat{X}_{FFT}(n,m) = \sum_{m=0}^{N-1} x(k + nN) w(k) e^{-j2\pi \frac{mk}{N}}
\end{equation}

where n, m, N, and w represent the frame index, the discrete-frequency index, the frame length, and the analysis window function, respectively. 
The polar form of the STFT of the noisy speech is represented by

\begin{equation}
\hat{X}_{FFT}(n,m) = \left| \hat{X}_{FFT}(n,m) \right| 
e^{j \angle \hat{X}_{FFT}(n,m)} = F(n,m)
\end{equation}

where F(n,m) is the phase value and A(n,m) is the magnitude value. To prepare the features of the noisy signal and the targets of clean speech as the input and output of the model, the logarithm of the magnitude A(n,m) is calculated.  These logarithmic magnitudes represent the log power spectrogram features of the mix data. It will be normalized to make the training process faster and fed to the model as the input. The structure of the model is explained in the following section.

\begin{itemize}
    \item Raw Audio Branch Pre-Processing
\end{itemize}

In addition to the spectral features, the proposed framework also incorporates the raw audio signal as a complementary input. In this branch, the noisy signal, X(k), is first filtered using Filter 1 to remove out-of-band frequencies. Then, instead of framing and overlapping as in the spectral branch, consecutive frames of the raw audio are first stacked without overlap and then averaged to produce a single representative feature vector for the segment.

This averaging step reduces the temporal redundancy while preserving the essential patterns of the raw waveform. The resulting vector, denoted as ${X}_{raw}(k)$ , captures the temporal characteristics of the audio signal in a compact form. This processed raw audio feature is normalized and combined with the spectral features in the model input pipeline to leverage both time-domain and frequency-domain information.

The combined input strategy enables the framework to better adapt to real-world conditions, where both spectral and temporal cues are crucial for differentiating speech from complex and dynamic noise sources. The structure of the transformer-based model that processes these inputs is explained in the following section.

\subsection{Model Structure}

The noise suppression model is built upon a modified Vision Transformer (ViT) architecture designed to combine multi-modal features and capture the complex spatiotemporal and frequency domain patterns in noisy signals. This dual-input architecture, illustrated in Figure 2, is composed of two primary branches, the spectrogram branch and the raw audio branch, which process distinct feature types to extract complementary information for robust noise suppression.

\begin{figure}[t]
  \centering
  \includegraphics[width=0.7\linewidth]{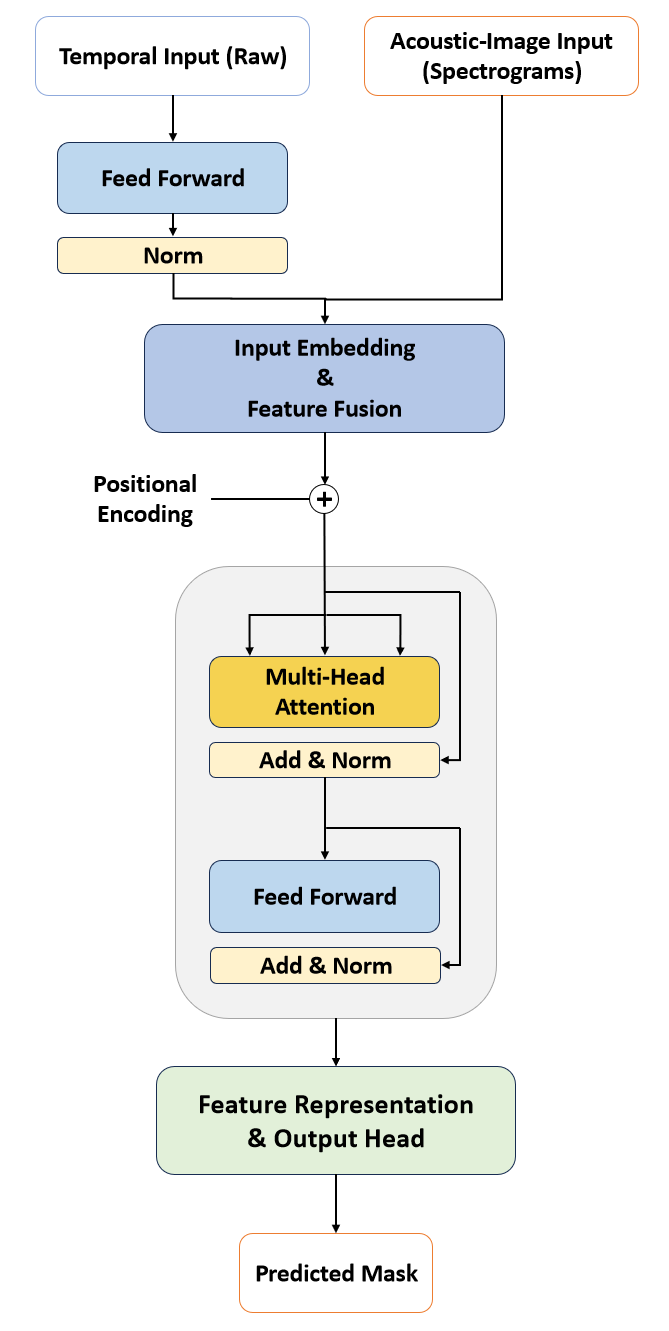}
   \caption{The model structure.}
   \label{fig:onecol}
\end{figure}

\begin{itemize}
    \item Spectrogram Branch
\end{itemize}
This branch processes normalized magnitude spectrogram features, as described in the preprocessing section. These features provide the high-resolution frequency content essential for visual pattern recognition by the ViT core. The output of this branch is reshaped into a sequence of visual tokens ready for the transformer block.

\begin{itemize}
    \item Raw Audio Branch
\end{itemize}

The raw audio branch processes the waveform features directly, capturing time-domain information that is not explicitly available in the frequency domain. This branch is designed to be computationally lightweight and includes the following steps:

\begin{enumerate}
    \item Input Transformation: The processed audio features are normalized and transformed into a higher-dimensional dense embedding using a fully connected layer.

    \item Normalization and Reshaping: The embedding undergoes batch normalization to stabilize optimization and is then reshaped into a compatible format for concatenation. This process ensures that the temporal features are correctly integrated with the spectral tokens.
\end{enumerate}

The outputs from the spectrogram and raw audio branches are concatenated (Feature Fusion) and processed through a transformer-based core. Key components of this core include:

\begin{itemize}
    \item Multi-Head Attention (MAH): Each MAH layer captures long-range dependencies not only between the spectral features (tokens) but also between the spectral and the raw audio features. This mechanism helps the model focus on relevant information while ignoring irrelevant noise components.
\end{itemize}

\begin{itemize}
    \item Residual Connections: Standard residual connections are implemented alongside layer normalization to maintain efficient information flow and prevent gradient degradation across deep layers.
\end{itemize}

The proposed model employs a lightweight architecture modified to fuse spectral and raw audio representations through two transformer layers with four attention heads. 

Following the transformer blocks, the learned feature representation undergoes flattening and is passed through the final Multilayer Perceptron (MLP) head for regression. This head uses several fully connected layers with ReLU activations and Dropout to refine features before the final Output Layer, which produces the predicted mask for audio reconstruction. The network weights are updated using the ADAM optimizer, which uses adaptive learning rates for efficient gradient descent optimization ~\cite{kingma2014adam}.

\subsection{Dual Mask Prediction}

The model predicts two types of masks:

Clean mask $M_{FFT}(k)$: This mask is computed using the energy of clean and noisy signals and is used to extract the clean signal components. The mask is computed using the following equation,

\begin{equation}
M_{FFT}(k) = \left( \frac{S^2(n,m)}{S^2(n,m) + X^2(n,m)} \right)^{0.5}
\end{equation}

where the ${S^2(n,m)}$ and $X^2(n,m)$ are the energy of the clean signal and mixed signals at the ${n^{th}}$ frame and ${m^{th}}$ frequency bin, respectively ~\cite{rizwan2020snmf}.

Noise Mask $M_{wav}(k)$: This mask is derived using the energy of noise and noisy signals to extract the noise components.

\begin{equation}
M_{wav}(k) = \left( \frac{N^2(n,m)}{N^2(n,m) + X^2(n,m)} \right)^{0.5}
\end{equation}

The predicted masks are applied to the input noisy signal to reconstruct both the clean and the noise components, enabling a comprehensive noise suppression framework.

\subsection{Post-processing}

Post-processing steps, as indicated in Figure 1, include the predicted mask modification, reconstruction, and tuning of the clean signal in the time domain. Instead of using the raw output of the model, the two mask vectors (Clean mask and Noise mask) are passed through filter 2, a Gaussian smoothing function widely used for denoising in image and audio processing. The level of smoothing is controlled by the standard deviation of the Gaussian function. The smoothed masks are then used in the following steps to create both the clean speech signal and the noise signal.

For the clean signal, the Clean mask is multiplied by the magnitude of the mixed signal to obtain the clean magnitude, as described by the equation:

\begin{equation}
P(k) = \hat{M}_{FFT}(k)^{\frac{2}{L}} A(k)
\end{equation}

Similarly, the Noise mask is multiplied by the mixed signal’s magnitude to extract the noise signal. The noise magnitude is computed as:

\begin{equation}
N(k) = \hat{M}_{wav}(k)^{\frac{2}{L}} A(k)
\end{equation}

The final magnitude of the clean signal is the difference between the clean and noise signals:

\begin{equation}
S(k) = P(k) - N(k)
\end{equation}

Up to this point, the magnitude of the clean signal is computed in the frequency domain. To transform the signal to the time domain, the IFFT tool (Inverse FFT) is used. During the process, the clean signal is reconstructed by vectorizing the complex part of the computed magnitude and the phase of the noisy mixture. This step is completed in the Complex box in Figure 1, using equations 9 and 10.

In the end, the predicted clean speech is passed through filter 3, which is a bandpass filter for a final tune to improve the quality of the predicted signal.

\begin{equation}
\hat{S(k)} = F(k) + jP(k)
\end{equation}
\begin{equation}
The predicted clean speech = Inverse STFT(\hat{S(k)}) 
\end{equation}

\subsection{Experimental Setup}

In this section, the performance of the proposed architecture is examined using the Librispeech ~\cite{panayotov2015librispeech} and Urbansound8K ~\cite{salamon2014urbansound}, and GoogleAudioset ~\cite{gemmeke2017audioset} datasets. 900 utterances from various English speakers (female and male) are randomly selected from the first dataset as the clean speech for training and testing purposes. For the background sounds, various recordings were selected to simulate different environments such as homes, restaurants, and residential neighborhoods. In addition, the UrbanSound8K and Google Audioset datasets, which include both stationary noises (e.g., engine sounds) and non-stationary noises (e.g., dog barking and sirens), were employed as the noise dataset. The noise samples are selected such that they include a clear noise over the entire audio sample. Clean speech and noise signals are single-channel and down-sampled to 8kHz to adjust the dataset size for training and also speed up the real-time processing practice. As mentioned in Section 2.1, the selected noise and clean signals are added together to build the mixed signal, which is used as the input to the model. The effect of SNR variation is examined in this section. In these experiments, the performance of the trained model is evaluated with four different SNRs, which include 0 dB, 3 dB, 10 dB, and -3 dB, to simulate the real-world situation. It should be noted that 20\% of the created training set is selected for validation and testing.

\subsection{ Real-time Processing}
In order to make the algorithm work in real-time, the frame size is chosen such that the human auditory system cannot realize the delay caused by pre-processing, noise suppression, and post-processing. Human ears can not recognize a 20-40 ms latency, and the proposed framework can keep the latency in this range to work smoothly in real-time. To meet this requirement, a frame length of 16 ms was chosen as an optimal trade-off between temporal and frequency resolution while maintaining computational efficiency. This duration corresponds to 128 samples at an 8 kHz sampling rate, a power-of-two length that allows fast FFT computation and minimizes latency without compromising spectral detail.

The overall sound latency of the deployed system was verified using an external measurement protocol. This involved using two synchronized microphones: one positioned to capture the input sound and the other to capture the final filtered output sound. The latency was determined by comparing the time difference between the waveforms captured by these two devices. The inference time for the model and processing blocks was measured directly using code profiling tools.

\section{Results and Discussion}

In this section, the simulation results of the proposed technique are presented. Several experiments under the same test conditions are performed to evaluate the behaviour and the performance of the proposed method under various and complex noisy situations. To evaluate the efficiency of the proposed algorithm, four standard and most commonly used speech quality measurements are used, including:

\begin{itemize}
    \item Perceptual Evaluation Speech Quality (PESQ):  is the most common metric to evaluate speech quality, calculated by comparing the enhanced and clean speech signals. PESQ is calculated as a linear combination of average disturbance values and average asymmetrical disturbance values between a reference signal and an estimated signal. Although it assesses the noise speech quality, some features such as loudness, loss, delay, and echo are not expressed in the PESQ score ~\cite{gogate2020cochleanet}. In the PESQ test, the representative values are from -0.5 to 4.5, demonstrating the minimum and maximum speech quality. 
\end{itemize}

\begin{itemize}
    \item Short-Time Objective Intelligibility (STOI): reflects the improvement in speech intelligibility and has a strong correlation to subjective listening test scores. STOI is based on an intermediate intelligibility measure for the short-time Time-Frequency domain and uses a simple DFT-based TF-decomposition ~\cite{taal2010stoi}. STOI score ranges from 0 to 1, where a better speech intelligibility results in a higher STOI score. 
\end{itemize}

\begin{itemize}
    \item Segmental Signal to Noise Ratio (Seg SNR): evaluates the overall speech quality and the performance of noise reduction. Seg SNR computes the segmental signal-to-noise ratio (SNR) in dB by comparing a noisy signal with a clean reference signal. The Seg SNR for a speech signal is formulated as follows,

    \begin{equation}
    SegSNR = mean\left(10 \log_{10} 
    \left( 
    \frac{\sum_{n=1}^{N} x^2(n)}
    {\sum_{n=1}^{N} (x(n) - \hat{x}(n))^2}
    \right)
    \right)
    \end{equation}

    where N is the length of the signal frame, x(n)  is the clean speech signal, and $\hat{x}(n)$ is the enhanced speech signal ~\cite{huang2020multiband}.
\end{itemize}

\begin{itemize}
    \item The Log-Likelihood Ratio (LLR): represents the quality level of speech signals by measuring the difference between the linear prediction of the clean reference signal and the degraded signal ~\cite{hu2008objective}. LLR score ranges from 0 to 2, where higher LLR scores mean better speech qualities. 
\end{itemize}

Table 1 presents the experimental results of the average performance based on PESQ, STOI, Seg SNR, and LLR scores on the selected test samples. For the test samples, four different unseen audios from each noise type are selected, which include highly dynamic noises (siren), non-stationary noises (vacuum cleaner), and stationary noises (engine). Dog barking is also considered the fourth sample to show the algorithm’s performance in a difficult situation where the noise has structural similarity to the speech (from the spectrograms representing clean speech and dog bark in Figure 3). The signals are mixed with four different SNRs, i.e., -3 dB, 0 dB, 3 dB, and 10 dB, to include various noisy conditions. For all these measurements, higher scores indicate better performances. According to Table 1, the simulation results show that the proposed method provides high speech intelligibility with significant speech quality improvements. The results indicate that the enhanced speech quality is the highest in 10dB and is degraded as it gets to -3dB. This is due to the noise level amplification. Moreover, the results demonstrate that the proposed framework has significantly enhanced speech quality even at higher noise levels of 3dB and 0dB.

\begin{table}[t]
  \caption{The performance measurements of the proposed method for different noise structure types.}
  \label{tab:performance}
  \centering
  \begin{tabular}{lccccc}
    \toprule
    \textbf{ } & \textbf{SNRs} & \textbf{PESQ} & \textbf{STOI} & \textbf{Seg SNR} & \textbf{LLR} \\
    \midrule
    \multicolumn{6}{l}{\textbf{Siren}} \\
      & 0 dB  & 2.3564 & 0.8504 & 3.1065 & 1.5719 \\
      & 3 dB  & 2.8372 & 0.8680 & 3.7530 & 1.6983 \\
      & 10 dB & 3.2567 & 0.8980 & 3.9237 & 1.8601 \\
      & -3 dB & 1.9162 & 0.7160 & 1.2519 & 1.2749 \\
    \midrule
    \multicolumn{6}{l}{\textbf{Engine}} \\
      & 0 dB  & 2.4095 & 0.9788 & 4.2853 & 1.6011 \\
      & 3 dB  & 2.6525 & 0.9848 & 4.7296 & 1.6791 \\
      & 10 dB & 3.1156 & 0.8163 & 4.8516 & 1.8164 \\
      & -3 dB & 2.0196 & 0.7026 & 2.5174 & 1.3725 \\
    \midrule
    \multicolumn{6}{l}{\textbf{Vacuum Cleaner}} \\
      & 0 dB  & 2.5080 & 0.8804 & 3.2365 & 1.5697 \\
      & 3 dB  & 2.7352 & 0.8941 & 4.0709 & 1.5738 \\
      & 10 dB & 3.4297 & 0.9116 & 4.8583 & 1.5764 \\
      & -3 dB & 2.1998 & 0.7576 & 1.9869 & 1.4671 \\
    \midrule
    \multicolumn{6}{l}{\textbf{Dog Barking}} \\
      & 0 dB  & 2.4782 & 0.8844 & 3.8941 & 1.4813 \\
      & 3 dB  & 2.7153 & 0.8891 & 4.1668 & 1.5095 \\
      & 10 dB & 3.1299 & 0.9164 & 4.8946 & 1.5818 \\
      & -3 dB & 2.0148 & 0.7644 & 1.4928 & 1.3941 \\
    \bottomrule
  \end{tabular}
\end{table}

\begin{figure*}
  \centering
  \includegraphics[width=0.7\linewidth]{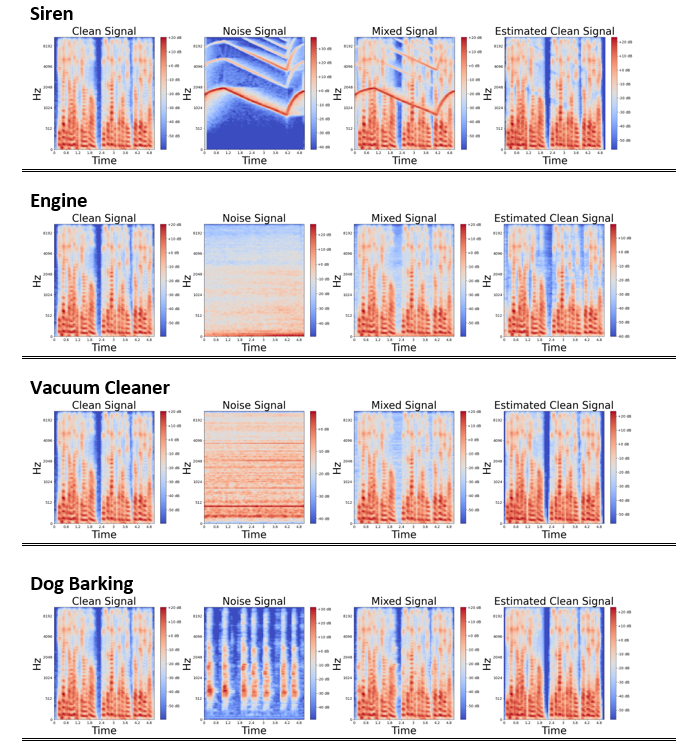}
  \caption{Spectrogram performance of the proposed method for different noise structure types.}
\end{figure*}

To visually observe the experimental results, the spectrograms of the clean speech signal, the noise signal, the mixture of the speech signal with the noise at an SNR of 0 dB, and the enhanced speech estimated by our proposed method are presented in Figure 3. In a spectrogram, the vertical axis indicates the frequency values in Hertz, the horizontal axis shows time steps, and the color brightness represents amplitude values. According to Figure 3, it is observed that the proposed method effectively suppresses the background noise, and the enhanced speech signal is close to the clean version. In highly dynamic noise structures such as dog barking and sirens, eliminating the background noise without any corruption to the speech signal is complicated and difficult. Figure 3 shows that our proposed method is capable of suppressing the highly dynamic noises, and the resulting filtered speech is close to the clean speech spectrogram. 

To benchmark the proposed noise suppression system against established speech enhancement techniques, we compared its performance with an LSTM network ~\cite{westhausen2020dltn}, a widely recognized real-time baseline in the literature. Table 2 summarizes the results in terms of PESQ, STOI, and SI-SDR. The Scale-Invariant Signal-to-Distortion Ratio (SI-SDR) measures how accurately the enhanced signal preserves the target speech structure while remaining independent of overall signal scale. Unlike SegSNR, which evaluates local frame-level signal-to-noise ratios, SI-SDR focuses on distortion between the reference and estimated signals after optimal scaling. The proposed system achieved PESQ values ranging from 1.9 to 3.32, STOI between 0.70 and 0.98, and SI-SDR improvements of 6 to 15 dB across our test samples. These results are comparable to or better than those of ~\cite{westhausen2020dltn} (PESQ 2.23 – 3.04, STOI 0.80 – 0.95, SI-SDR 2 – 16 dB), despite operating at a lower sampling rate (8 kHz vs. 16 kHz) and being evaluated on more challenging real-world noises such as sirens, dog barking, vacuum cleaning, and engine sounds.

\begin{table*}[t]
  \caption{An objective comparison between the proposed noise suppression model and ~\cite{westhausen2020dltn}.}
  \label{tab:comparison}
  \centering
  \begin{tabular}{l l c c c c}
    \toprule
    \textbf{Method} & \textbf{Dataset} & \textbf{Sampling Rate} & \textbf{PESQ} & \textbf{SI-SDR} & \textbf {STOI} \\
    \midrule
    \textbf{LSTM model [31]} & WHAMR / DNS Challenge & 16 kHz & 2.23--3.04 & 2--16 & 0.80--0.95 \\
    \textbf{Proposed System} & LibriSpeech + GoogleAudioSet & 8 kHz & 1.9--3.42 & 6--15 & 0.70--0.98 \\
    \bottomrule
  \end{tabular}
\end{table*}

While ~\cite{westhausen2020dltn} and related benchmarks are typically trained and tested on stationary or semi-stationary noises from datasets like DNS Challenge or WHAMR, our system demonstrates strong performance under highly dynamic, non-stationary noise conditions. This highlights the robustness and generalization capability of the proposed model for realistic environmental applications.

\begin{figure}[t]
  \centering
  \includegraphics[width=0.8\linewidth]{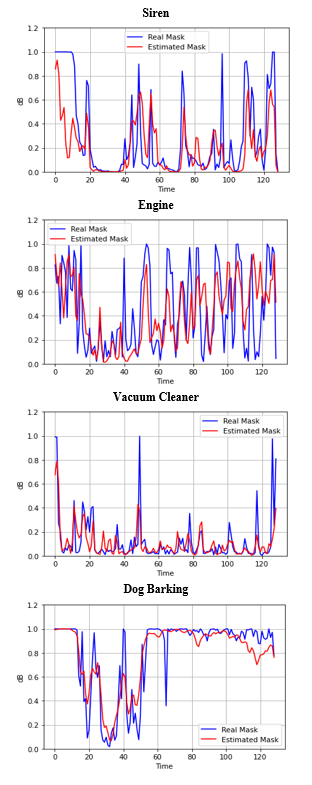}
   \caption{A comparison between the estimated masks by the proposed method and the real masks.}
   \label{fig:onecol}
\end{figure}

To further discuss the effectiveness and capability of our method, the estimated clean mask, which is one of the outputs of the ViT model, is compared with the computed clean mask in equation 6, which is considered the real clean mask. Figure 4 shows four randomly selected samples of the estimated clean masks compared to the real ones computed directly from the clean reference and the mixed noisy signals. The simulation results show that the predicted real masks are very close to each other, which indicates the accuracy and performance of the proposed system. According to Figure 4, the model can predict the fluctuations in the ratio masks and generate a clean speech at the end of the process.

The end-to-end latency of the system running on an NVIDIA Jetson Xavier NX GPU is measured. The total latency per frame was observed to be under 35 ms, which is comfortably within the acceptable human auditory range. Table 3 shows the measured latency breakdown for a single 16 ms  frame:

\begin{table}[t]
  \caption{Latency breakdown for a single 16 ms frame.}
  \label{tab:latency}
  \centering
  \begin{tabular}{lc}
    \toprule
    \textbf{Components} & \textbf{Latency (ms)} \\
    \midrule
    Frame Length & $\sim$16 \\
    Model Inference & $\sim$7 \\
    Pre- and Post-Processing & $\sim$3 to 4 \\
    Operating System Overhead & $\sim$8 \\
    \midrule
    \textbf{Total End-to-End Latency} & $\sim$34 to 35 \\
    \bottomrule
  \end{tabular}
\end{table}

\section{Conclusion}

This paper proposes an innovative solution to address the single-channel noise suppression problem using a transformer-based learning framework for real-time applications. The proposed framework receives a noisy speech signal as an input and extracts its features to feed to a ViT model. The model is designed and trained to generate two ratio masks to eliminate the background noise from the noisy signal. Signal processing techniques are utilized to increase speech quality and integrability before and after the model. The proposed techniques and model are applied on the Librispeech dataset for the clean speech sources and the GoogleAudioset and UrbanSound8K datasets for the noise sources. The simulation results indicate that the proposed method can remove the background noise from the speech signal, even in the presence of complex noisy conditions and difficult noise types. Four speech quality measurements, including PESQ, STOI, Seg SNR, and LLR, are employed to evaluate the performance of the proposed method in terms of the noise reduction level, speech quality, and intelligibility. Moreover, the spectrograms and the predicted masks are provided to demonstrate the significant performance of the proposed technique under different circumstances, from low to high SNRs.

{
    \small
    \bibliographystyle{IEEEtran}
    \bibliography{main}
}


\end{document}